\documentclass[copyright,creativecommons]{eptcs}
\providecommand{\event}{TARK 2025}

\usepackage{iftex}
\usepackage[T1]{fontenc}
\ifpdf\usepackage{underscore}\else\usepackage{breakurl}\fi

\usepackage{libertine} \usepackage[ruled,vlined]{algorithm2e}

\SetAlgorithmName{Program}{}{}
\newenvironment{program}[1][ht]
  {\begin{algorithm}[#1]
  }{\end{algorithm}}

\usepackage{amsmath,amsthm,amssymb}
\usepackage[bibliography=common,appendix=strip]{apxproof}
\usepackage{hyperref}

\newtheorem{definition}{Definition}
\newtheoremrep{theorem}{Theorem}
\newtheoremrep{lemma}{Lemma}
\newtheoremrep{corollary}{Corollary}
\newtheoremrep{proposition}{Proposition}

\usepackage{color} 
\newcommand{\Agents}{\mathit{Agt}}
\newcommand{\Values}{\mathit{Values}}
\newcommand{\diff}{\mathit{diff}}
\newcommand{\exchange}{\mathcal{E}}
\newcommand{\I}{\mathcal{I}}
\newcommand{\N}{\mathcal{N}}

\newcommand{\R}{\mathcal{R}}
\newcommand{\W}{\mathcal{W}}
\newcommand{\F}{\mathcal{F}}
\newcommand{\crashed}{\mathit{crashed}}
\newcommand{\clean}{\mathit{clean}}
\newcommand{\noop}{\mathtt{noop}}
\newcommand{\decide}{\mathtt{decide}}
\newcommand{\flood}{\mathit{flood}}
\newcommand{\counting}{\mathit{count}}
\newcommand{\kbp}{\mathbf{P}}
\newcommand{\opt}{\textsc{opt}}
\newcommand{\send}{\mathit{send}}
\newcommand{\Time}{\mathit{time}}

\newcommand{\Crash}{\mathit{Crash}}

\newcommand{\init}{\mathit{init}}
\newcommand{\Nat}{\mathbb{N}}

\DeclareMathOperator{\deciding}{deciding}

\begin{document}

\title{Optimality of Simultaneous Consensus with Limited Information Exchange (Extended Abstract)\footnote{This research is supported by an Australian Government Research Training Program (RTP) Scholarship. Research supported by AFOSR grant FA23862114029. The Commonwealth of Australia (represented by the Defence Science and Technology Group) supported this research through a Defence Science Partnerships agreement.}}

\author{Kaya Alpturer
\institute{Princeton University\\
New Jersey, USA}
\email{kalpturer@princeton.edu}
\and
Ron van der Meyden
\institute{School of Computer Science and Engineering\\
UNSW Sydney\\
Sydney, Australia}
\email{r.vandermeyden@unsw.edu.au}
\and
Sushmita Ruj
\institute{School of Computer Science and Engineering\\
UNSW Sydney\\
Sydney, Australia}
\email{sushmita.ruj@unsw.edu.au}
\and
Godfrey Wong
\institute{School of Computer Science and Engineering\\
UNSW Sydney\\
Sydney, Australia}
\email{godfrey.wong@student.unsw.edu.au}
}

\def\titlerunning{Optimality of Simultaneous Consensus with Limited Information Exchange}
\def\authorrunning{K. Alpturer, R. van der Meyden, S. Ruj and G. Wong}

\maketitle
\begin{abstract}
Work on the development of optimal
fault-tolerant
Agreement protocols using the logic of knowledge has concentrated on the ``full information'' approach to information exchange, which is costly with respect to message size. 
Alpturer, Halpern, and van der Meyden 
(PODC 2023)
introduced the notion
of optimality with respect to a limited information exchange, 
and studied the 
\emph{Eventual}
Agreement problem
in the sending omissions failure model.
The present paper studies the 
\emph{Simultaneous}
Agreement problem 
for the crash failures model, and a number of limited information exchanges from the literature. 
In particular, the paper considers information exchanges from a 
FloodSet protocol (Lynch, Distributed Algorithms 1996), 
a variant of this in which agents also count the number of failures (Casta{\~{n}}eda et al, NETYS 2017), and a variant 
in which agents associate each agent with a value (Raynal, PRDC 2002). 
A new information exchange is also introduced that enables decisions to be made at worst one round later than the optimal protocol of Dwork and Moses (I\&C 88), but with lower computation cost and space requirements.
By determining implementations of a knowledge based program, 
protocols are derived that are optimal 
amongst protocols for each of these information exchanges.
\end{abstract}

\begingroup
\let\clearpage\relax
\section{Introduction}

Epistemic logic 
has been shown to play a useful role 
in the analysis of
distributed algorithms \cite{HM90}.
By characterizing the knowledge that an agent needs to have 
in order to take an action, 
it is possible to obtain distributed algorithms that are optimal in the sense of terminating
as soon as any other algorithm that solves the same problem.

In particular, variants of the Byzantine Agreement
problem have been fruitfully studied using this methodology. 
(The term Byzantine agreement is used in literature even when
the failure model is  
weaker than Byzantine failures
\cite{DM90,MT88,FHMVbook,AHM23}.)
In Byzantine Agreement, a group of agents is required to 
make 
a
common decision, in a setting where there could be faulty agents in the system that might be
unreliable or even malicious.
In the Simultaneous 
Byzantine Agreement (SBA)
version of this problem, nonfaulty agents are required to make the decision at the same time (in the same round of computation).
Reaching simultaneous agreement was shown to be related to
common knowledge \cite{DM90}, and this characterization 
has been used to derive optimal protocols for 
a variety of failure models, including crash \cite{DM90}
and omission failures \cite{MT88}.

In the present paper, we reconsider this type of optimality result in the particular case of crash failures and a synchronous message passing communications model. We assume synchronized clocks and a known upper bound on message delivery time, so that the computation can be structured into rounds. 
In a distributed system prone to crash failures,
a faulty agent can crash during a round and
send an arbitrary subset of the 
messages it was supposed to send.
In the rounds following the crash,
the crashed agent no longer sends any messages.

In the quest for optimality, most existing literature on using epistemic logic to study
distributed algorithms 
has 
worked with  \emph{full information protocols}, in which 
each agent broadcasts its complete local state in each round, 
and stores all messages received. This results in states
that grow exponentially with time. Even when the state size
can be optimized, there are still cases where the computation
required to be performed in each round is intractable. For example, in the general omissions model, optimal implementations require
$NP$-hard computations \cite{MT88, Moses09}.
This means that practical protocols will often exchange less information than does the full information protocol.
This raises the following question: having made a compromise on the information that is exchanged by the agents, is the protocol making full use of \emph{that} information. That is, is the protocol
optimal amongst protocols that use the same pattern of information exchange, terminating no later than does any other such protocol?

To date, only one 
paper has studied 
optimality of
Byzantine agreement protocols with respect to limited information exchange \cite{AHM23}.
There, 
the authors studied the
\emph{eventual} Byzantine agreement problem in
synchronous message passing systems
that are prone to sending omission failures.
In the present paper, we will study the 
Simultaneous Byzantine Agreement (SBA) 
problem in synchronous message passing systems
that are prone to crash failures.

We study a number of limited information exchanges derived from protocols in the literature, as well as one new information exchange ``SendWaste''. 
The main contribution of 
the 
paper is
to derive protocols that are optimal with respect to these information exchanges, by identifying, in each run, the earliest time that agents are able to make a decision. 
In a number of cases, this analysis establishes decision times that are earlier than the times stated in the source publications, showing that the protocols in the literature can be optimized.

For FloodSet \cite{Lynch96}, 
in which the information exchanged in each round is the set of initial values an agent has heard about,
we are able to terminate as soon as
$m \ge \min\{t+1,n-1\}$, where $m$ is time,
$t$ is maximum possible number of failures in a run,
and $n$ is the total number of agents.
(When $t=n-1$, this is earlier than the time $t+1$ usually stated.)

For Counting FloodSet \cite{CMRR17}, which is an extension of FloodSet
that counts the number of messages an agent has received,
the 
earliest 
stopping condition
is the same as FloodSet, except for the case
when an agent does not receive any message from other agents.

\begin{table}[ht]
\begin{center}
\begin{tabular}{|l|lll|} \hline
Protocol & Computation & Message size & Space \\ \hline
FloodSet \cite{Lynch96} & $O(n)$ & $O(1)$ &
$O(1)$ \\
Counting FloodSet \cite{CMRR17} & $O(n)$ & $O(1)$ &
$O(1)$ \\
Vectorized FloodSet \cite{Raynal02} & $O(n^2)$ & $O(n)$ & $O(n)$ \\
SendWaste 
[this paper]
& $O(n)$ & $O(1)$ & 
$O(1)$ \\
Dwork and Moses \cite{DM90} & $O(nt)$ & $O(t)$ & 
$O(n)$ \\
\hline
\end{tabular}\vspace{2mm}
\caption{Summary of computational complexity per round,
message size, and space complexity for each agent.
}
\label{table:complexity}
\end{center}
\end{table}

We 
also show 
that even by extending Counting FloodSet
to store a history of number of missing messages in each round,
we are unable to obtain 
a
stopping condition
that terminates earlier.
This protocol is of interest because in EBA,
an earlier stopping condition can be obtained by retaining the count of the 2 previous rounds.
We show that for SBA, there is no such advantage.
However, we pay a price of more space used, for no advantage in the stopping time. So, while this protocol is optimal with respect to its own information exchange, it is still possible to improve this protocol by reducing its space usage. 

For 
another variant, Vectorized FloodSet \cite{Raynal02}, in which
agents 
keep not just a 
set
of initial values, but also know which agents have those initial values, 
and in which a nonfaulty agent sends a message only when it learns a new 
agent-value pair, 
we 
establish an interesting 
nontrivial early stopping condition, that improves upon the condition $m=t+1$ given in \cite{Raynal02}. 
In Vectorized FloodSet,
we can decide as soon as the time is greater than
$\min\{t+1,n-1\} -\max\{1, \beta_i (r,m)\}$,
where $\beta_i (r,m)$ is the number of messages
not received by agent $i$ during round 1 of run $r$.

The most interesting protocol we 
consider 
is 
a new protocol SendWaste that we introduce. 
We 
show that it can terminate no later than one round after
the protocol
of Dwork and Moses \cite{DM90},
which is optimal with respect to full-information exchange,
but at lower computation cost and space usage, particularly when the number of possible failures $t$ is large. 
The protocol of Dwork and Moses maintains several sets of agents (one of which it transmits), from which it computes a notion of ``Waste''. Our new protocol SendWaste estimates the waste and transmits the estimate. This reduces both message size and computation costs.

The computational complexity per round, message size,
and space complexity for each agent
for the protocols we have studied
is summarised in \autoref{table:complexity}. (For the protocols that send or store numerical values, we assume that the number $n$ can be stored in a single machine word, and that arithmetic operations have cost 1, since we expect this to be the case for practically feasible systems. Beyond this, some logarithmic adjustments would be required.)

Among the protocols studied in this paper,
FloodSet performs the worst in every run,
which terminates in $\min\{t+1,n-1\}$ rounds.
Counting FloodSet has the same optimal early stopping condition
except for the case where an agent detects $n-1$ missing messages.
Counting FloodSet with Perfect Recall has the same
optimal early stopping condition as Counting FloodSet in every run.
Vectorized FloodSet is able to terminate as soon as Counting FloodSet
except for the case where an agent 
does not receive $n-1$ messages. 
In this case, Counting FloodSet is able to decide
while Vectorized FloodSet is unable to decide.
SendWaste is able to terminate as early as any other
FloodSet variants studied in this paper.
The protocol of Dwork and Moses \cite{DM90} is
optimal with respect to full-information exchange and hence it is
at least as good as any other protocol for any information exchange.

We 
begin by 
introducing our model of the problem
in \autoref{sec:framework}.
Then, in \autoref{sec:optimality} we discuss
what it means for a Simultaneous Byzantine Agreement protocol
to be optimal
relative to an information exchange protocol.
We give an analysis of 
the
FloodSet protocol in \autoref{sec:floodset}
and its counting variants in \autoref{sec:counting}.
\autoref{sec:raynal} is an analysis of 
the Vectorized FloodSet 
protocol from \cite{Raynal02}. 
Our new protocol SendWaste is treated in Section~\ref{sec:send}.
Finally, we conclude this paper in \autoref{conclusion}.

 \section{Framework} \label{sec:framework}
We will mostly follow the framework from
\cite{AHM23} and \cite{Meyden24}.
In the Byzantine agreement problem, 
agents from a set 
$\Agents = \{1, \dots, n\}$
communicate using a synchronous, round-based, message-passing network
in order to decide on a value. 
Each agent starts with an initial value from the set 
$\Values=\{0,1\}$.
We assume that $t < n$ is the maximum number of failures 
that can occur in a run. We focus on \emph{crash} failures, 
in which a faulty agent $j$ acts correctly until a crash occurs
in some round $m$ in which an arbitrary subset of its messages in round 
$m$ are delivered. Moreover, agent $j$ never sends a message again 
in rounds $m' > m$.

The specification of 
Eventual 
Byzantine Agreement (EBA) can be stated as follows:
\begin{itemize}
\item \textbf{Agreement:} 
if nonfailed agents $i$ and $j$ decide on values $v$ and $v'$, respectively, then $v=v'$
\item \textbf{Validity:} 
if a nonfailed agent decides on value v, then some agent has $v$ as its initial value.
\end{itemize}
We remark that, in Agreement, we could have that $i=j$, and the times at which a decision is made could be different. Thus, the specification implies that once an agent has made a decision, it never makes a different decision. We allow that an agent makes the same decision at different times. 
In this paper, we do not adopt the ``Unique Decision'' requirement that an agent makes a decision at most once. 
The specification of the Simultaneous Byzantine Agreement (SBA) problem
involves all the rules of Eventual Byzantine Agreement problem
together with simultaneity: 
\begin{itemize}
\item \textbf{Simultaneity:} 
if a nonfailed agent decides in a round, then every nonfailed agent decides in the same round.
\end{itemize}
Various formulations of a termination condition have been considered in the literature. The protocols we study satisfy 
\begin{itemize} 
\item \textbf{Termination:} every agent eventually decides or fails,
\end{itemize}

We decompose SBA protocols into two components $(P,\exchange)$, comprised of a decision protocol $P$ and 
an information exchange protocol $\exchange$.  
We represent the states of the environment in which agents operate by a set $L_e$, 
defined more precisely below.
An information exchange protocol describes 
what information the agents record in their local states, and 
what messages
an agent sends in a given local state.
Formally, an 
information exchange protocol is a tuple 
$\langle \exchange_1, \ldots, \exchange_n \rangle$
containing a local information exchange protocol $\exchange_i$  for each agent $i$. 
A local information exchange protocol $\exchange_i$ of agent $i$ is a tuple
$\langle L_i, I_i, A_i, M_i, \mu_i, \delta_i \rangle$ that consists~of:
\begin{itemize}
\item a set of local states $L_i$,
\item a set of initial states $I_i \subseteq L_i$,
\item a set of allowed actions $A_i$ for agent $i$
\item a set $M_i$ of messages that are allowed to be sent by agent $i$,
\item a message selection function $\mu_i: L_i \times A_i \to \prod_{j \in \Agents} (M_i \cup \{\bot\})$,
\item a transition function
$\delta_i: L_i \times L_e \times A_i \times \prod_{j \in \Agents} (M_j \cup \{\bot\}) \to L_i$.
\end{itemize}
For SBA protocols, we assume that there exist, for each agent $i$, a function $\init_i:I_i \rightarrow \Values$ that identifies an agent's initial value, 
and a function $\Time_i:L_i \rightarrow \Nat$ that identifies the time on the agent's clock. 
We assume that the function $\init_i$ is a bijection, that is, that there is a unique initial state for each value in $\Values$.
The message selection function $\mu_i$ takes a local state and an action
performed by an agent in a round 
to a tuple of messages that 
the agent sends in that round. 
The transition function $\delta_i$ updates the local state depending
on 
the state of the environment, 
the action an agent performs in the round and the set of messages it receives in that round.

In the full information exchange, the local initial states
are the initial values of the agent.
In every round, 
each agent sends 
their local state to every other agent,
and records every 
message 
it receives in its local state.

A decision protocol 
describes what actions the agents perform in a given situation.
For the SBA problem, the action set for each agent $i$ is $A_i = \{\decide_i(v)~|~v\in \Values\} \cup \{\noop\}$.
A \emph{local decision protocol} $P_i: L_i \to A_i$ for agent $i$
maps a local state 
of agent $i$ 
to an action that $i$ can perform.
A \emph{decision protocol} $P$ is a tuple $\langle P_1, \ldots, P_n \rangle$
of local decision protocols for all agents.

It will help when comparing two information exchanges below to assume that agent's decisions do not affect the information that they transmit. 
We say that an information exchange $\exchange$ \emph{does not 
transmit information about actions}
if, for all agents $i$, local states $s\in L_i$, and actions $a_1,a_2 \in A_i$, 
\begin{itemize}
\item $\mu_i(s,a_1) = \mu_i(s,a_2)$ (that is, the messages the agent sends do not depend on their action), and  
\item for all message vectors $\mathtt{m}$ and environment state $s_e \in L_e$, we have $\delta_i(s,s_e,a_1,\mathtt{m}) = \delta_i(s,s_e,a_2,\mathtt{m})$ (that is, the agent's state update does not depend on their action).  
\end{itemize}
The definition does not allow an agent to record the particular decision it has made in its local state, since it could then transmit information about this decision in the next step, since $\mu_i$ still depends on the local state. 
All the concrete information exchanges we consider in this paper have 
the property that they do not transmit information about actions. 

A failure model describes the failures that could occur
in a run of a protocol.
We use the hard crash failure model 
$\Crash_t$, depending on a parameter $t$ that is a number less than the number of agents.
In 
this 
failure model, 
there may be up to $t$ faulty agents, who 
can crash at any time.
During the round in which an agent 
crashes, 
it sends a subset of 
the 
messages it is supposed to send.
After an agent has crashed 
it does not send any message
and no longer runs the protocol.

We represent 
$\Crash_t$ as a set of \emph{adversaries}, where an adversary is a function
$\F: \mathbb{N} \times \Agents \times \Agents \rightarrow \{\top,\bot\}$.
If agent $i$ crashes in round $m$, then we  require that $F(m,i,i) = \bot$. 
If the message that agent $i$ 
was supposed to send to agent $j$ is not sent in round $m$, 
we have $\F(m,i,j)=\bot$, otherwise $\F(m,i,j)=\top$.
If 
$\F(m,i,i)=\bot$ 
then $\F(m',i,k)=\bot$ for 
all times 
$m'>m$ and every $k \in \Agents$.
There may be at most $t$ agents $i$ such that $\F(m,i,j) = \bot$ for some $m$ and $j$.
We take the set of states of the environment $L_e$ to be the set of all such adversaries. 
Furthermore, for this model, we assume that the set of local states of agent $i$ contains a special
state $\crashed$, representing that the agent has crashed.
The state update function operates so that in a round $m$ in which the adversary $s_e = \F$ and 
$\F(m,i,i) = \bot$,
we have $\delta_i(s_i,s_e,a,v) = \crashed$ for all 
local states $s_i$ of agent $i$, 
actions $a$ and vectors $v$ of messages.

A global state 
is an element of 
$L_e \times \prod_{i \in \Agents} L_i$
consisting of a tuple comprised of the 
state of the environment
and a local state
of each agent.
A \emph{run} is a mapping $r: \mathbb{N} \to L_e \times \prod_{j \in \Agents} L_j$
from times to a global states.
If $r(m) = \langle s_e, s_1, \ldots ,s_n\rangle$, we write 
$r_e(m)$ for $s_e$ and, for each agent $i$,  $r_i(m)$ for $s_i$.
The time between $(r,m)$ and $(r,m+1)$ is \emph{round} $m+1$.
A pair $(r,m)$ comprised of a run $r$ and a time $m$ is a \emph{point}.

A \emph{system} is a set $\mathcal{R}$ of runs. 
An \emph{interpreted system} is a pair $\I = (\R,\pi)$, 
where $\R$ is a system and $\pi$ is an \emph{intepretation} 
mapping each point $(r,m)$ of $\I$ to a function $\pi(r,m): \mathit{Prop} \to \{true,false\}$
that determines whether or not each atomic proposition from a set $\mathit{Prop}$ 
is true at the point.

Given an information exchange $\exchange$ and decision protocol $P$ for the crash failure model 
$\Crash_t$, we construct an interpreted system 
$\I_{P, \exchange,\Crash_t} 
= (\R,\pi)$
as follows. 
(Generally, $\Crash_t$ will be implicit, and we write $\I_{P, \exchange}$ for brevity.)
The runs in $\R$ are generated by selecting an initial global state $r(0) =
\langle s_e, s_1, \ldots, s_n \rangle$, where
$s_e \in L_e = Crash_t$ is an environment state encoding an 
adversary from the crash failure model and $s_i \in I_i$ for all $i \in \Agents$ are initial local states of agents. The remainder of the run is uniquely determined from the initial global state by the 
following induction. 
The state of the environment $r_e(m)$ will be the same for all times $m$. 
For each round $k+1$, the local state $r_i (k+1)$ of agent $i$ is determined 
as follows. First, each agent $i$ uses the decision protocol $P_i$ to select its action $a_i = P_i(r_i(k))$, 
and 
attempts to send 
the messages $\mu_i(r_i(m),a_i)$. The adversary $r_e(k)$ determines
which of these messages are delivered, so that each agent $i$ receives an agent indexed vector $v_i$
of messages (or $\bot$ in case of a message that was not sent or is not delivered
because the agent sending it crashed). 
Agent $i$'s local state $r_i (k+1)$ is then equal to $\delta_i(r_i(k), r_e(k), a_i,v_i)$.
We consider only \emph{synchronous} protocols, for which 
$\Time_i(r_i(m)) =m$ for all points $(r,m)$ of runs of the protocol.

Note that if $\exchange$ does not 
transmit 
information about actions, then the set of runs of $\I_{P, \exchange}$ does not depend on $P$. 
We may therefore write simply  $\I_{\exchange}$
when not referring to propositions about the protocol.

The atomic propositions 
$\mathit{Prop}$ in the interpretation $\pi$
consist of:
\begin{itemize}
\item $\exists v$, which is true if an agent in the current run
has $v$ as its initial value,
\item $i\in \N$, which is true if agent $i$ has not failed up to the current time, and 
\item $\clean$, which is true if a clean round has occurred
(defined below).
\end{itemize}

We work with a modal logic that 
uses the atomic propositions along with the
standard boolean operators $\lnot, \lor, \land$. 
For an agent $i$ and 
an \emph{indexical} set $\N$ of agents (which may depend on the point at which we evaluate the formula)
we also have
unary 
modal operators 
$K_i, E_{\N}, C_{\N}$
which we will define later. 
The semantics of the logic in an interpreted system $\I$ is given by a relation $\models$, such that $(\I,r,m) \models \phi$,
for a point $(r,m)$ of $\I$ and a formula $\phi$, represents that $\phi$ is true at the point $(r,m)$ of $\I$. 
The definition of this relation is given by an induction on the construction of the formula $\phi$. 
For an atomic proposition $p$,  we have $(\mathcal I,r,m) \models p$, if $\pi(r,m)(p) = \textbf{True}$. 

The semantics of the modal operators is given using 
an indistinguishability relation $\sim_i$ on points for each agent $i$. The points
$(r,m)$ and $(r',m')$ are indistinguishable to agent $i$,
written as $(r,m) \sim_i (r',m')$, if $r_i(m) = r_i'(m')$.
Note that the indistinguishability relation is an equivalence relation.
The intuition behind the definition of knowledge is that
agent $i$ knows $\varphi$ if $\varphi$ is true in all
the points that are indistinguishable to $i$.

For a 
formula $\varphi$ and point $(r,m)$
of an interpreted system $\I$,
we say:
\begin{itemize}
\item agent $i$ knows $\varphi$ at $(r,m)$,
written as $(\mathcal I,r,m) \models K_i \varphi$, if
for all points $(r',m')$ of $\I$ satisfying $(r,m) \sim_i (r',m')$,
we have $(\mathcal I,r',m') \models \varphi$,
\item everyone in 
$\N$
knows $\varphi$ at $(r,m)$,
written as $(\I,r,m) \models 
E_\N
\varphi$, if
for every 
$i \in \N(r,m)$,
we have $(\I,r,m) \models K_i (\varphi)$,
\item $\varphi$ is common knowledge among 
$\N$
at the point
$(r,m)$,
written as 
$(\I,r,m) \models C_\N \varphi$, if
for every $k \in \mathbb Z^+$, we have $(M,s) \models E_\N^k \varphi$.
\item $\varphi$ is distributed knowledge 
amongst group $\N$,
written as $(\I,r,m) \models 
D_\N \varphi$,
if $(\I,r',m) \models \varphi$ for all \\
$(r',m) \in\bigcap_{i \in \N(r,m)} S_i (r,m)$, where 
$S_i (r,m)$ is the set of points of $\I$ that are
indistinguishable to agent $i$ from $(r,m)$.
\end{itemize}
A formula $\varphi$ is valid in $\mathcal I$,
written as $\mathcal I \models \varphi$,
if for every point $(r,m)$ of $\mathcal I$,
we have $(\mathcal I,r,m) \models \varphi$.

In this paper, 
the indexical set $\N$ represents
the set of nonfailed agents. 
Formally, $i \in \N(r,m)$ iff $r_i(m) \neq \crashed$.

\section{Optimality and Knowledge-based Programs}\label{sec:optimality}
In this section we study knowledge-based programs
and what it means for a 
protocol
to be optimal
with respect to a given information exchange.

To compare two protocols, we consider their behaviour in runs in which the environment has the same behaviour. 
Two runs $r \in \I$ and $r' \in \I'$ 
\emph{correspond} 
if they have identical initial values and the failure pattern is the same in both runs,
that is, for all agents $i$, we have $\init_i(r_i(0))= \init_i(r'_i(0))$ and  $r_e(0) = r'_e(0)$.

To define optimality, we define a partial order on protocols. 
Let $P$ and $P'$ be two decision protocols for the same information exchange  $\exchange$. 
We write $P \leq_\exchange P'$ if for all corresponding runs $r$ of $P$ and $r'$ of $P'$ with respect to $\exchange$, and for all times $m$, if $P_i(r_i(m)) = \decide_i(v)$ for some value $v$, 
then $P'_i(r'_i(m)) = \decide_i(w)$ for some value $w$. That is, at any time when $P$ makes a decision, $P'$ also makes a decision (but possibly a different one). We define $P$ to be an 
\emph{optimal SBA protocol with respect to $\exchange$} if $P$ is an SBA protocol with respect to $\exchange$, and for all SBA protocols $P'$ with respect to $\exchange$, if $P \leq_\exchange P'$ then $P' \leq_\exchange P$. That is, there does not exist an SBA protocol $P'$ with respect to 
$\exchange$ that makes decisions strictly more often than $P$. 
Protocol $P$ is an \emph{optimum SBA protocol with respect to $\exchange$} if it is an SBA protocol with respect to $\exchange$ 
such that for all SBA protocols $P'$ with respect to $\exchange$, 
we have $P' \leq_\exchange P$. 

These definitions of optimality are related to other definitions of optimality in the literature
\cite{AHM23,CGM14}, which focus on optimizing the earliest time at which a decision is made, 
as follows. Note that if $P \leq_\exchange P'$, then protocol $P$ never makes its first decision earlier than $P'$ makes its first decision. Thus, an optimal SBA protocol cannot be improved with respect to the earliest time at which it makes a decision on any run, and an optimum protocol always decides as least as early as any other protocol.

The following 
lemma is taken from Moses and Tuttle \cite{MT88}.
Its proof does not assume a full information protocol.
The proposition $\deciding_i (v)$ is defined to be true at a point $(r,m)$
if agent $i$ decides $v$ in round $m+1$ of run $r$, i.e., $P_i(r_i(m)) = \decide_i(v)$.
\begin{lemma}\cite{MT88} \label{lem:MT88-lem4}
Let $r$ be a run of 
an SBA protocol generating interpreted system $\I$ 
and 
let $i\in \N(r,m)$. 
If $(\I,r,m) \models \deciding_i(v)$ for $v \in \{0,1\}$, then
$(\I,r,m) \models
C_{\N} (\exists v)$.
\end{lemma}

It is useful to know that a failure-free run
requires at least $\min\{t+1,n-1\}$ rounds
to attain common knowledge of initial values.
Therefore, when time $m < \min\{t+1,n-1\}$
and $r$ is a run of any
SBA 
protocol,
if $(r,m)$ is $\N$-reachable from a failure-free run
then we do not 
have common knowledge of any initial value at $(r,m)$.

\begin{lemma} \label{lem:FIP decision time} \cite{DM90}
Let $r$ be a failure-free run of
an
SBA protocol generating system $\I$. 
When $m<\min\{t+1,n-1\}$ we have
$(\I,r,m) \models \lnot C_{\N} (\exists 0)
\land \lnot C_{\N} (\exists 1)$.
\end{lemma}

Moses and Tuttle \cite{MT88} have proven
that a full information protocol attains common knowledge no later than any arbitrary SBA protocol.
A \emph{proposition $\varphi$ is about the environment} if its truth value in 
interpreted systems for the SBA problem, as defined above, 
depends only on
the adversary and the  initial values of the agents. That is, for corresponding runs $r$ and $r'$ of systems $\I$ and $\I'$, respectively, 
and all times $m$, we have $(\I,r,m) \models \varphi$ iff $(\I',r',m) \models \varphi$.

\begin{lemma} \cite[Corollary 6]{MT88} \label{lem:MT88 FIP best}
Let $\varphi$ be a proposition about the 
environment.
Let $r$ and $r'$ be corresponding runs of an arbitrary protocol
$(P,\exchange)$ 
and a full-information protocol 
$(P',\exchange')$
respectively.
If $(
\I_{P,\exchange}
,r,m) \models C_{\N} \varphi$,
then $(
\I_{P',\exchange'}
,r',m) \models C_{\N} \varphi$.
\end{lemma}

Let $\exchange^1$ and $\exchange^2$ be two information exchange protocols that do not 
transmit information about actions. 
We say $\exchange^1$ \emph{stores at least as much information as} $\exchange^2$ 
if, for every 
pair of 
corresponding runs $r^1$ and $r^2$
of 
$\I_{\exchange_1}$ and $\I_{\exchange_2}$, respectively, 
and every 
pair of 
corresponding runs $r^3$ and $r^4$
of $\I_{\exchange_1}$ and $\I_{\exchange_2}$, respectively, 
$(r^1,m) \sim_i (r^3,m)$ implies $(r^2,m) \sim_i (r^4,m)$.
This relation is reflexive and transitive.

\begin{lemmarep} \label{lem:reachable info}
Let 
$\exchange^1$ be an information exchange that stores at least as much information as information
exchange
$\exchange^2$ 
and let $r^1$ and $r^2$ be corresponding runs
of $\I_{\exchange^1}$ and $\I_{\exchange^2}$ respectively.
If $r^3$ and $r^4$ are corresponding runs,
of $\I_{\exchange^1}$ and $\I_{\exchange^2}$ 
 respectively, such that
$(r^3,m)$ is $\N$-reachable from $(r^1,m)$
in $k$ steps,
then
$(r^4,m)$ is $\N$-reachable from $(r^2,m)$
in $k$ steps.
\end{lemmarep}

We now get the following generalization of \autoref{lem:MT88 FIP best}.

\begin{propositionrep} \label{thm:less info no CK}
Suppose $\varphi$ is a proposition about the environment.
Suppose $\exchange^1$  stores at least as much information as protocol $\exchange^2$ 
and let $r^1$ and $r^2$ be corresponding runs
of $\I_{\exchange^1}$ and $\I_{\exchange^2}$, respectively. 
Then
$(\I_{\exchange^1},r^1,m) \models
\lnot C_{\N} \varphi$ implies
$(\I_{\exchange^2},r^2,m) \models
\lnot C_{\N} \varphi$.
\end{propositionrep}

We remark that 
\autoref{thm:less info no CK} 
does 
not make assumptions about the failure model and
would work even with arbitrary failures.

\emph{Knowledge based programs} \cite{FHMVbook} 
can be understood as specifications, in a code-like form, that describe how each agent's actions relate to 
its knowledge. Syntactically, they are like standard programs for an agent, except that the formulas that occur in 
as the conditions of conditional statements may be a formula of the logic of knowledge stating a property 
of the agent's knowledge, rather than just a predicate of the agent's local state, as in standard programs. 
Program~\ref{kbp} gives an example, which will be the specific knowledge based program that we study in the present paper. 
This program is to be interpreted as the rule used to determine agent $i$'s action in each round, until one
of the actions $\decide_i(v)$ is selected, after which the program terminates.
In this program, $\exists 0$ and $\exists 1$ means
there exists an agent with initial value of 0 and 1 respectively.

\begin{program}
    \DontPrintSemicolon
    \lIf{$K_i(C_{\N} (\exists 0))$}{$\decide_i(0)$}
    \lElseIf{$K_i(C_{\N} (\exists 1))$}{$\decide_i(1)$}
    \lElse{$\noop$}
    \caption{$\kbp^\opt_i$}
    \label{kbp}
\end{program}

The semantics of knowledge based programs is given by relating concrete 
protocols to a knowledge based program. Since we only consider one specific
program in this paper, we do not give this definition in general, and just explain this relation 
for the specific program 
$\kbp^\opt_i$.
Intuitively, a decision protocol $P$ implements 
the knowledge based program with respect to an information exchange $\exchange$, 
if, in each round, 
it selects for each agent $i$ 
the same action as would be obtained for agent $i$ from the rule in the knowledge based program. 
``If-then-else" statements are interpreted according to the usual semantics, but to interpret the knowledge formulas, 
we need an interpreted system. For this, we  use the interpreted system obtained
from running $P$ with respect to $\exchange$. 

Formally, given an interpreted system $\I$, for each agent $i$ we define the decision protocol $(\kbp^\opt_i)^\I$ 
on a local state $s\in L_i$ by taking $(\kbp^\opt_i)^\I(s)$ to be the action $a$ selected  the rule $\kbp^\opt_i$, 
with the truth values of the knowledge formulas evaluated at any point $(r,m)$ of $\I$ with $r_i(m) = s$. 
Note that the truth values of these formulas are independent of the choice of such an $(r,m)$, since they 
depend only on the local state of agent $i$ at that point. 
Using this, we define a decision protocol \emph{$P$ to be an implementation of Program  $\kbp^\opt$ with respect to $\exchange$ in the crash failure context $\Crash_t$} if 
for all points $(r,m)$ of $\I_{\exchange,\Crash_t}$  and all agents $i\in \N(r,m)$, 
we have $(\kbp^\opt_I)^\I(r_i(m)) = P_i(r_i(m))$. 

To obtain SBA protocols, we use a special case of the Agreement  
property: 
\begin{itemize} 
\item[] {\bf Self-Agreement:} For all agents $i$, if $i$ decides value $v$ and later 
decides value $v'$, then $v=v'$.
\end{itemize}
The following result shows that implementations of the knowledge based program 
$\kbp^\opt$ satisfying this property are SBA protocols.

\begin{theoremrep}
Suppose $P$ is an implementation of $\kbp^\opt$ with respect to information exchange 
$\Crash_t$ and $P$ satisfies Self-Agreement. 
Then $P$ is an SBA protocol, that is, 
 $\I_{P, \exchange, \Crash_t}$ satisfies SBA. 
\end{theoremrep}

The following result shows that implementations of the knowledge based program $\kbp^\opt$ with respect to an information exchange 
yield SBA protocols that are 
optimum
with respect to that information exchange. In the following sections, we apply this result by considering 
a number of information exchanges and deriving protocols that are implementations of $\kbp^\opt$ with respect to these  information exchanges. 
The resulting protocols will therefore be 
optimum
SBA protocols with respect to their information exchanges.

\begin{theoremrep} \label{thm:optimal-sba}
    Let an SBA protocol $P$ be an implementation of~~$\kbp^\opt$ 
with respect to information exchange $\exchange$ in the crash failure context $\Crash_t$.
Suppose that $\exchange$ does not transmit information about actions.
Then, $P$ is an optimum SBA protocol
with respect to $\exchange$ in the crash failure context $\Crash_t$.
\end{theoremrep}

Using \autoref{thm:optimal-sba}, we can show 
an SBA protocol to be an optimum SBA protocol 
by showing it is an implementation of $\kbp^\opt$.
In the following sections, We do this for a number of concrete protocols. In particular, we consider 
FloodSet in \autoref{sec:floodset},
Counting FloodSet in \autoref{sec:counting},
Vectorized FloodSet in \autoref{sec:raynal},
and SendWaste in \autoref{sec:send}.
 \section{FloodSet} \label{sec:floodset}
Lynch introduced the FloodSet protocol \cite{Lynch96}
as an example of a simple protocol that solves
SBA under crash failures.
In this section, we present a refined analysis of the Floodset 
protocol in terms of the agents' knowledge, showing that
after a minor modification, the 
Floodset protocol is 
an optimum SBA protocol,
given the amount of information the agents are exchanging.

We now define the FloodSet protocol in two parts using our framework, 
the information-exchange protocol $\exchange^\flood$ 
and the decision protocol $P^\flood$.
The local states of agent $i$ in $\exchange^\flood$ will be 
either the special state $\crashed_i$, or
a tuple
$\langle W_i, m, v_i \rangle$, where
$W_i$ is a set of distinct values agent $i$ has seen so far,
$m$ is the current time,
and $v_i$ is the initial value of $i$.
The information-exchange protocol $\exchange^\flood$
has the set of initial local states $I_i$ for each agent $i$ consisting of all states of the form
$\langle \{v_i\}, 0, v_i \rangle$
for some $v_i\in \Values$.
In each round, every agent $i$ sends $W_i$ to all other agents.
That is, $\mu_i((W_i,m,v_i), a_i) = (W_i, \ldots, W_i)$ for all $(W_i,m,v_i)\in L_i$ and actions $a_i$, 
and $\mu_i(\crashed_i,a_i) = (\bot, \ldots, \bot)$. 
At the end of a round
$m+1$, 
$v_i$ stays the same, 
and the local state of each agent is updated to have a new 
$W_i$ which is the union of all messages received in the current round.
That is, the function $\delta_i$ is defined by 
$$\delta_i\left(\left(W_i,m,v_i\right), \F,a_i, \left(W_1',\ldots,W_n'\right)\right) = \left(W_i \cup \bigcup_{j} W_j', m+1,v_i\right)$$
provided the environment does not crash agent $i$, i.e., $\F(m,i,i) \neq \bot$, otherwise 
$$\delta_i((W_i,m,v_i), \F,a_i, (W_1',\ldots,W_n')) = \crashed_i$$
and $\delta_i(\crashed_i, \F,a_i, (W_1',\ldots,W_n')) = \crashed_i$.

Throughout this section, we fix $t \leq n$ and write $\I$ for $\I_{\exchange_{\flood},\Crash_t}$.
When agent $i$'s local state at point $(r,m)$ of $\I$ is $r_i(m) = (W,m,v)$, we define $W_i(r,m) $ to be $W$, and 
$\Time_i(r,m)$ to be $m$.

The standard program 
from \cite{Lynch96}
representing the FloodSet decision protocol 
for an agent $i$ such that $P^\flood = (P^\flood_1, \dots, P^\flood_n)$ 
is
Program~\ref{alg:lynch}.

\begin{program}[ht]
    \DontPrintSemicolon
    \lIf{$m = t+1$}{$\decide_i(\min W_i)$}
    \lElse{ $\noop$ }
    \label{alg:lynch}
    \caption{$P^\flood_i$ from \cite{Lynch96}}
\end{program}

A crucial definition used in the analysis 
of SBA protocols 
is the notion of 
a \emph{clean round}, 
in which every nonfailed agent
receives the same messages.
\begin{definition}[Clean Round]
A round is clean if for any nonfailed agents $i$ and $j$,
agent $i$ receiving a message from agent $k$ implies that
agent $j$ also received a message from agent $k$.
\end{definition}

The following
show the importance of clean rounds.
\begin{lemma} \cite[Lemma 6.1]{Lynch96} \label{lem:lynch1}
If round $m$ of run $r$ of FloodSet is clean, then
$W_i (r,m) = W_j (r,m)$ for all nonfailed agents $i,j$
in run $r$.
\end{lemma}
\begin{lemma} \cite[Lemma 6.2]{Lynch96} \label{lem:lynch2}
Suppose $W_i (r,m) = W_j (r,m)$ for any nonfailed agents
$i,j$ in run $r$ of FloodSet.
Then, $W_i (r,m') = W_j (r,m')$ for all $m \le m' \le t+1$.
\end{lemma}

Therefore, since we can only have up to $t$ crashes, by 
round $t+1$ we are guaranteed to have a clean round.  
Moreover, all agents $i \in \Agents$ have the same $W_i(r,t+1)$
in any run $r$ in 
$\I$. 
Note that however, when $t = n-1$, 
either we have a clean round by time $t$, or exactly one 
agent failed in every round until round $t$, 
so 
that only one 
active agent is remaining. In both cases, we should be able to decide
without waiting until $t+1$.
This observation leads to the following refinement to the Floodset decision protocol.
Let $P^{\flood+} = (P^{\flood+}_1,\dots,P^{\flood+}_n)$
be the decision protocol defined by the following standard program:

\begin{program}
    \DontPrintSemicolon
    \lIf{$m \ge \min\{t+1,n-1\}$}{$\decide_i(\min W_i)$}
    \lElse{ $\noop$ }
    \label{alg:lynch+}
    \caption{$P^{\flood+}_i$}
\end{program}

\begin{theorem}
Let $(r,m)$ be a point in $\I$. 
We have $(\I, r, m) \models \bigvee_{v \in \Values} K_i(C_{\N}(\exists v))$ for all $i \in \N(r,m)$ if and only if $m \ge \min\{t+1, n-1\}$.
\end{theorem}

By Theorem~\ref{thm:optimal-sba}, this result implies
that the protocol $P^{\flood+}$ is optimal with 
respect to the information-exchange $\exchange^\flood$. 
 \section{Counting Extensions} \label{sec:counting}
The aim of this section is to explore what happens
if an agent stores the number of missing messages
during a round.
We aim to find out whether we can achieve a better
early stopping condition by making the above change
to the FloodSet information-exchange protocol $\exchange^\flood$.
This modification was studied in \cite{CMRR17}
and was shown to 
lead to an 
early stopping condition
for EBA.
However, there is no literature on this modification
for SBA.
Furthermore, we will extend Counting FloodSet to
store the entire history of number of missing messages
to investigate if we are able to obtain an
earlier stopping condition. 
We consider this variant because  \cite{CMRR17} shows that a predicate based on 
the \emph{difference} between two successive counts of missing messages leads to an EBA protocol that 
stops earlier than the protocol based just on a single count. Again, we  are interested in whether this information is 
helpful in the setting of SBA.

The local states of agent $i$ in the Counting FloodSet 
information-exchange protocol $\exchange^{\counting}$
will be either the state $\crashed_i$ or a tuple $\langle W_i, h_i, m, v_i \rangle$, where
$W_i$ is the set of distinct values agent $i$ has seen so far,
$h_i$ is the number of missing messages during round $m$,
$m$ is the current time,
and $v_i$ is the initial value of $i$.
The initial state of an agent $i$ which
starts with initial value of $v_i$
is 
$\langle \{v_i\}, 0, 0, v_i \rangle$.
As in the FloodSet protocol, in each round, each nonfailed agent $i$ sends a message comprised of the set $W_i$ to all other agents. 
The state update function adds any values in a message $W_j$ received from another agent to the set $W_i$, but additionally sets 
$h_i$ to the number of agents from which $i$ did not receive a message in the current round. The time $m$ is incremented in each round, 
but the value $v_i$ is unchanged. 

The local states of agent $i$ in 
the Counting FloodSet with perfect recall information-exchange protocol
$\exchange^{\counting(pr)}$
will be a tuple $\langle W_i, h_i, m, v_i \rangle$, where
$W_i$ is the set of distinct values agent $i$ has seen so far,
$h_i$ is a vector in which $h_i[k]$ is
the number of missing messages during round $k$,
$m$ is the current time,
and $v_i$ is the initial value of $i$.
The initial state of an agent $i$ which
starts with initial value of $v_i$
is 
$\langle \{v_i\}, [0], 0, v_i \rangle$.
Messages and state updates are as in the Counting FloodSet protocol, except that the count of 
missing messages is appended to $h_i$. 

The decision rules used in the early stopping condition in both information-exchanges can be found in the following programs:

\begin{program}
\DontPrintSemicolon
    \lIf{$m \ge \min\{t+1,n-1\}$}{decide$(\min W_i)$}
    \lElseIf{$h_i \geq n - 1$}{decide$(\min W_i)$}
    \lElse{ noop }
\label{alg:count}
\caption{Counting FloodSet decision rule}
\end{program}

\begin{program}
\DontPrintSemicolon
    \lIf{$m \ge \min\{t+1,n-1\}$}{$\decide_i(\min W_i)$}
    \lElseIf{$\exists k\leq m (h_i[k] \geq n - 1)$}{$\decide_i(\min W_i)$}
    \lElse{$\noop$}
\label{alg:perfect}
\caption{Counting FloodSet (perfect recall) decision rule}
\end{program}

\begin{lemmarep} \label{lem:floodset info order}
The following relations hold amongst the versions of the FloodSet information exchange:
\begin{enumerate}
\item $\exchange^{\counting(pr)}$ stores at least as much 
      information as $\exchange^{\counting}$.
\item $\exchange^{\counting}$ stores at least as much
      information as $\exchange^{\flood}$.
\end{enumerate}
\end{lemmarep}

The following results characterize the situations in which agents obtain common knowledge when using the counting variants of FloodSet. 
\begin{theorem}
Let $r$ be a run of the Counting FloodSet protocol,
let $\I = \I_{\exchange^{\counting}, \Crash_t}$, and let $m$ be a time.
When 
the local state of agent $i$ at point $(r,m)$ is $r_i (m) =\langle W, h, m, v_i \rangle$
we have $(\I,r,m) \models 
\bigvee_{v \in \Values}
C_{\N}(\exists v)$
if and only if
$m \ge \min\{t+1,n-1\}$ or $h \ge n-1$.
\end{theorem}
\begin{theorem}
Let $r$ be a run of Counting FloodSet with perfect recall,
let $\I = \I_{\exchange^{\counting(pr)}, \Crash_t}$, and let $m$ be a time.
When 
the local state of agent $i$ at point $(r,m)$ is $r_i (m) =\langle W, h, m, v_i \rangle$
we have $(\I,r,m) \models 
\bigvee_{v \in \Values}
C_{\N}(\exists v)$
if and only if
$m \ge \min\{t+1,n-1\}$ or there exists $1 \le k \le m$ such that $h[k] \ge n-1$,
\end{theorem}

We have proven that Counting FloodSet
and Counting FloodSet with perfect recall are
optimum SBA protocols 
with respect to their information exchange, due to 
Theorem~\ref{thm:optimal-sba}.
 \section{Vectorized FloodSet} \label{sec:raynal}
Raynal \cite{Raynal02} describes a protocol that 
differs from the FloodSet protocol
by recording 
not just the initial values received but 
also which agents had these initial values.
Instead of sending a message every round,
an agent 
only sends 
newly received information, 
immediately after
the round 
in which 
the agent receives the new information.
As in the Floodset protocol, agents do not 
learn 
whether 
any other agent
has detected a failure.
Agents also do 
not record the round number in which a message is received.
The local states 
of agent $i$ at a point
$(r,m)$ have the form
$\langle V_i(r,m), New_i(r,m), \Time_i(r,m) \rangle$
where $V_i (r,m)$ is an 
agent-indexed 
array of initial values
(or $\bot$, representing an unknown value),
$New_i (r,m)$ 
is a set of tuples of type 
$\Values\times \Agents$ 
recording 
new information that agent $i$ 
received in round $m$,
and 
$\Time_i(r,m) = m$
is the current time.
Intuitively, a tuple $(v,j) \in \Values\times Agents$ represents
the information that agent $j$ has initial preference $v$. 
This information is new to agent $i$ in round $k$ if 
at the start of the round (time $k-1$), the agent has $V_i[k](r,k-1) = \bot$
and the agent receives a message $New_{i'}$ in round $k$ containing the tuple 
$(v,j)$.  

Code for Vectorized FloodSet is given in the following figure. From this,
the information exchange protocol $\exchange^R$ for this 
protocol can be easily constructed. (We leave the details for the reader to complete.) 
For the remainder of this section, we write $\I$ for $\I_{\exchange^R,\Crash_t}$.

\begin{program}
\DontPrintSemicolon
$V_i \gets [\bot, \ldots, v_i, \ldots, \bot]$\\
$New_i \gets \{(v_i, i)\}$\\
\For {$r = 1, 2, \ldots, t+1$}{
begin round\\
\If {$New_i \ne \emptyset$}{
    \For {each agent $j \ne i$}
        {send $New_i$ to agent $j$}
}
$New_i \gets \emptyset$\\
$W_j \gets$ message received from agent $j$ or $\emptyset$ if none\\
\For {each agent $j \ne i$}{
    \For {each $(v,k) \in W_j$}{
        \If {$V_i [k] = \bot$}{
            $V_i[k] \gets v$\\
            $New_i \gets New_i \cup \{ (v,k) \}$
        }
    }
}
$W_j \gets \emptyset$\\
end round
}
\lIf { $\exists 0 \in V_i$ }{ decide 0 }
\lElse {decide 1} 
\label{alg:raynal}
\caption{Vectorized FloodSet for crash failures for agent $i$}
\end{program}

In Raynal's presentation, the protocol waits until the end of round $t+1$ before making a decision. We show that an optimal protocol using the same information exchange can make earlier decisions. Specifically, define $\beta_(r,m)$ be the number of $\bot$ values in the array $V_i(r,m)$. We show that the implementation of the knowledge based program uses the following predicate of agent $i$'s local state to determine the time $m$ at which a decision can be made:
$m > \min\{t+1, n-1\} -\max\{1,\beta_i (r,m)\}$.

The following theorem concludes when have 
common knowledge of initial values in Vectorized FloodSet.
\begin{theorem}
Let $r$ be any run in Vectorized FloodSet.
For any time $m$, we have $(\I, r, m) \models 
( C_{\N} (\exists 0) \lor C_{\N} (\exists 1) )$
if and only if
$m 
> 
 \min\{t+1, n-1\} -\max\{1, \beta_i (r,m)\}$.
\end{theorem}

We have found that Vectorized FloodSet is not optimal
with respect to its information exchange.
An
optimum 
early stopping condition is when
$m > \min\{t+1, n-1\} -\max\{1,\beta_i (r,m)\}$.

 \section{Sending the Waste} \label{sec:send}

Dwork and Moses 
\cite{DM90}
showed that an early 
decision
can be made at time
$m \ge \min\{t+1,n-1\} -\W(r)$, where
the waste $\W(r) = \max_{k \ge 0} \{ \diff (r,k)
\}$
and difference 
$\diff (r,k) =
\#KF(r,k) -k$
for 
$$\#KF(r,k) = \max_{0 \le \ell \le t}
\{\ell\ |\ (\I,r,k) \models D_{\N} (
\text{''$\ell$ agents have failed''}
) \}~.$$
They use this to develop a protocol
that is optimal with respect to the full information exchange. This protocol transmits a set of agents (newly detected to be faulty) in each round, and maintains a number of variables, each of type a set of (faulty) agents. To determine the set of agents from which it did not receive a message in a given round, the protocol requires a temporary variable recording the agents from which a message was received. This implies that the protocol has 
message size $O(t)$,
space requirement $O(n)$ and computation cost per round of $O(nt)$.

In this section, we develop a new protocol, that is able to make decisions almost as soon as the Dwork and Moses protocol, but at lower cost $O(1)$ in 
message size, 
and $O(n)$ in computation per round, 
and $O(1)$ in space costs. Whereas \cite{DM90} need to track which agents have been heard from, SendWaste need only \emph{count} the messages received in a round. 
The protocol is based on an information exchange in which 
agents maintain, 
in addition to a set of known values, 
a numerical estimate of the waste, which they transmit to other agents in each round. 
In this 
information exchange protocol, 
each agent 
estimates the waste 
at time $m$ 
by $h-m$,
where $h$ is the number of missing messages in 
round $m$.
The value $d$ stores the 
agent's 
estimate of the waste,
determined from the agent's own detection of missing messages and the 
waste estimates that it receives from other agents.

The local states of agent $i$ in the SendWaste
information-exchange protocol $\exchange^\send$
will be a tuple 
$\langle W_i, h_i, d_i, m, v_i \rangle$, where
$W_i$ is 
the 
set of distinct values agent $i$ has seen so far,
$h_i$ is 
the number of messages  missing from the last round,  
$d_i$ 
is the waste estimate,
$m$ is the current time, and
$v_i$ is the initial value of the agent.
The initial states are $\langle \{v\}, 0, 0, m, v \rangle$,
where $v \in \Values$.

In every round of SendWaste, agent $i$ sends 
$(W_i,d_i)$
to every agent.
Upon receiving messages
in round $m$,
agent $i$ updates $W_i$ to 
be the union of all $W_j$'s it receives and 
$d_i$ is updated to be 
the maximum of the agent's previous value of $d_i$, 
the values $d_j$ in messages $(W_j,d_j)$ received from 
other agents $j$, and $h_i-m$,
where 
$h_i$ is the number of missing messages in the current round.

For this information exchange, we define the SendWaste standard decision protocol $P^\send_i$
for an agent $i$ by 

\begin{program}
    \DontPrintSemicolon
    \lIf{$m \ge \min\{t+1,n-1\}-d_i$}{$\decide_i(\min W_i)$}
    \lElse{ $\noop$ }
    \label{alg:send}
    \caption{$P^\send_i$}
\end{program}
\noindent
and take $P^\send = (P^\send_1, \dots, P^\send_n)$. We will show that that $P^\send$ implements the knowledge based program $\mathbf{P}^{OPT}$ with respect to $\exchange^\send$, and is therefore an optimal SBA protocol with respect to that information exchange.

For the remainder of this section, we write $\I$ for $\I_{\exchange^\send,\Crash_t}$. 
We write  $d_{\N} (r,m)$ for the value $ 
\max_{i \in \N(r,m)} 
d_i (r,m)$.
The following theorem concludes when do we have and not have common knowledge of
initial values in SendWaste.
\begin{theorem}
Let $i$ be a nonfailed agent at $(r,m)$,
where $r$ is a run of SendWaste.
We have $(\I,r,m) \models \bigvee_{v \in \Values} C_{\N} (\exists v)$
if and only if $(\I,r,m) \models m \ge \min\{t+1,n-1\}-d_{\N}$.
\end{theorem}

We now show that SendWaste 
decides at most
one round later than the protocol from
Dwork and Moses \cite[Figure 3]{DM90},
which is 
an optimum 
with respect to full-information exchange.

\begin{theoremrep}
Let $r'$ be a run of Dwork and Moses' protocol \cite[Figure 3]{DM90}
and $r$ be its corresponding run of SendWaste.
If $r'$ decides at round $m'$, then $r$ decides at
some
$m$,
where $m' \le m \le m'+1$.
\end{theoremrep}

We have proven that SendWaste 
is
an optimum SBA protocol 
with respect to its information exchange.
In addition, the protocol is able to decide
at most one round later than a protocol that
is 
an optimum 
with respect to full-information exchange.
We can further improve the space complexity and message size
of SendWaste 
by only storing and sending $\min W$ instead of the whole set $W$.
With an $O(\log t)$
bit 
increase to message size
when compared to FloodSet, SendWaste 
has an
early stopping condition close to Dwork and Moses' protocol.
 \section{Conclusion} \label{conclusion}
We have studied the
FloodSet, Counting FloodSet, Counting FloodSet with perfect recall
and Vectorized FloodSet 
information exchanges 
for simultaneous Byzantine agreement
in synchronous message-passing systems prone to crash failures.
We have also introduced a new information exchange SendWaste in which agents transmit their known values and their estimate of the waste.
For each of these information exchanges,
we have determined the earliest
possible time that common knowledge about the initial values
is attained,
giving the earliest time at which agents are able to decide. 
This yields protocols that are 
an optimum 
for each of these information exchanges.

For FloodSet and its variants, the early stopping condition
is trivial.
On the other hand, Vectorized FloodSet has an
interesting nontrivial early stopping condition.

In addition, we came up with a limited information-exchange protocol
that decides at most one round later than
the Dwork and Moses \cite{DM90}
protocol that is 
optimum 
with respect to full-information exchange.
The new protocol, SendWaste,
has less computational complexity and space complexity 
than the 
Dwork and Moses
protocol.
The message size is also reduced
when compared to the protocol from Dwork and Moses.

We have used a version of the SBA specification in the present work that does not require that agents make a decision at most once, and have, indeed, allowed that agents continue to perform decision actions after they have first made a decision. Related to this, the local states in the concrete information exchange protocols we study do not record the fact that an agent has decided. This facilitates comparisons of protocols that make different decisions, or decide at different times. It also supports comparison of information exchange protocols with respect to the amount of information that they transmit, which facilitates transfer of conclusions from one protocol to another in our proofs. 

A downside of these assumptions is that our protocols 
do not satisfy the Unique Decision property that is 
often taken to be part of the SBA specification. 
We conjecture that, were we to modify these protocols 
so that they decide at most once, 
by adding a record of the fact that an agent has decided, then we would still be able to show that these protocols are the optimum SBA protocol for the given information exchanges, in the sense that they decide at least as early as all other SBA protocols for these information exchanges. 
Results of van der Meyden \cite{Meyden24} show that a terminating variant of the 
knowledge based program of the present paper yields \emph{optimum} SBA protocols in this sense, 
for information exchanges that do not transmit any information about actions, but may record decision information in a separate component of the local state. 
To apply this to our present protocols requires some
different techniques to show that the decision protocols are implementations of the knowledge based program. We leave this for future work.

In this paper, we considered the problem only under crash failures.
In future work, we intend 
to study the problem under
general omission failures. 
For this failure model, a polynomial time Simultaneous Byzantine Agreement  protocol, optimal with respect to its own information exchange, would be particularly interesting,
because an optimal protocol
with respect to the full information exchange is known to 
require NP-hard computations
in each round \cite{MT88}.
 \endgroup

\bibliographystyle{eptcs}
\bibliography{references}
\end{document}